\documentclass[aps,preprint,amsmath,amssymb]{revtex4}
\usepackage{graphicx}
\def\be{\begin{eqnarray}}
\def\en{\end{eqnarray}}
\def\non{\nonumber}

\def\tm{\tilde{m}}
\def\vpl{\vec{p}_{\ell}}
\def\vpk{\vec{p}_K}
\def\vpks{\vec{p}_{K^*}}
\begin{document}
\title{\Large \bf Probe the effects of split SUSY in rare $B$ decays}

\date{\today}
\author{\large \bf  Chuan-Hung~Chen$^{1,3}$\footnote{Email:
physchen@mail.ncku.edu.tw} and Chao-Qiang~Geng$^{2,3}$
\footnote{Email: geng@phys.nthu.edu.tw}   }
 \affiliation{$^{1}$Department of Physics,
National Cheng-Kung University, Tainan, 701
Taiwan\\
$^{2}$Department of Physics, National Tsing-Hua University, Hsin-Chu
, 300 Taiwan\\
 $^{3}$National Center for Theoretical Sciences, Taiwan }

\begin{abstract}
We study the decays of $B\to K^{(*)} \ell^{+} \ell^{-}$ in split
supersymmetry with R-parity violation. We find that the decay
branching ratio of $B\to K\tau^+\tau^-$ in the new physics model
due to the scalar interactions can be $1.8\times 10^{-6}$ which is
about one order of magnitude larger that in the standard model,
whereas those of $B\to K\ell^+\ell^-$ ($\ell=e$ and $\mu$) and the
$K^*$ modes are insensitive to the new physics. On the other hand,
the forward-backward asymmetries of $B\to K \tau^{+} \tau^{-}$ and
$K\mu^{+}\mu^{-}$, vanishing in the standard model, can be over
$10$ and  $1\%$,  respectively. In addition, we show that the new
interactions will significantly change the forward-backward
asymmetry in $B\to K^* \tau^{+} \tau^{-}$.
\end{abstract}
\maketitle

\section{Introduction}

One of the possible extension of the standard model (SM) is
supersymmetry (SUSY). It is found that the effects of SUSY at the
scale $\Lambda$ of $O(\rm TeV)$ can solve not only the hierarchy
problem, but also the problem of unified gauge coupling
\cite{unify1,unify2}. Moreover, the predicted lightest neutralino in
supersymmetric  models could also provide the candidate of dark
matter \cite{unify1,dark}.
In spite of the above successes, models with SUSY still suffer
some difficulties from phenomenological reasons, such as the
problems on small CP violating phases, large flavor mixings and
proton decays, as well as they predict too large cosmological
constant. Inevitably, fine tuning always appears in the low energy
physics. Recently, in order to explain the cosmological constant
problem and preserve the beauty of the ordinary low-energy SUSY
models, the scenario of split SUSY is suggested \cite{AD,ADGR}, in
which the SUSY breaking scale is much higher than the electroweak
scale. In this split SUSY scenario, except the SM Higgs which
could be as light as the current experimental limit, the scalar
particles are all ultra-heavy, denoted by $m_{S}\sim {\cal
O}(10^{9}-10^{13})$. On the other hand, by the protection of
approximate chiral symmetries, the masses of sfermions, such as
gauginos and higgsinos, could be at the electroweak scale
\cite{AD,GR}. 

In Ref \cite{CG-hepph},
we have found that due to
the large mixings of sneutrinos and the SM-like Higgs, the interesting
phenomena on the low energy $B_{s}$ system, such as the decays of
$B_{s}\to \ell^{+} \ell^{-}$ and the $B_{s}-\bar{B}_{s}$ mixing, could
occur in split SUSY R-parity violating models.
It is clear that
the same mixings
could also give some interesting implications on $B_{q}$
systems with $q=u$ and $d$.
In this paper, we discuss the possibility to probe the
effects of split SUSY in $B\to K^{(*)} \ell^{+} \ell^{-}$ decays.

The flavor-changing neutral current (FCNC) processes of $B\to
K^{(*)} \ell^+ \ell^-\ (\ell=e,\mu,\tau)$ are suppressed and
induced by electroweak penguin and box diagrams in the
SM with decay branching ratios (BRs) of ${\cal
O}(10^{-7}-10^{-6})$ \cite{Ali,MNS,CG-PRD}. The decay modes of
$B\to K^{(*)} \ell^+ \ell^-\ (\ell=e,\mu)$ have been  observed by
BELLE \cite{Bellebsll} with $Br(B\to K \ell^+
\ell^-)=(5.50^{+0.75}_{-0.70}\pm0.27\pm 0.02)\times 10^{-7}$ and
$Br(B\to K^{*} \ell^+ \ell^-)=(16.5^{+2.3}_{-2.2}\pm0.9\pm
0.4)\times 10^{-7}$, and by BABAR \cite{BaBarbsll} with $Br(B\to K
\ell^+ \ell^-)=(6.5^{+1.4}_{-1.3}\pm 0.4)\times 10^{-7}$ and
$Br(B\to K^{*} \ell^+ \ell^-)=(8.8^{+3.3}_{-2.9}\pm 1.0)\times
10^{-7}$. In addition, BELLE has also reported the
forward-backward asymmetry (FBA) as a function of dilepton
invariant mass in $B\to K^* \ell^{+} \ell^{-}$ \cite{Bellebsll}.
 Since the inclusive process
of $b\to s\ell^{+} \ell^{-}$ arises from loop corrections, these
exclusive FCNC rare decays are important for not only testing the SM
but also probing new physics. For instance, by considering polarized
leptons, various polarization asymmetries have been proposed
\cite{lep}. On the other hand, in terms of the $K^*$ polarizations
$\epsilon_{K^*}$, we can define a triple product operator
$\vec{\epsilon}_{K^*} \cdot (\vpks \times \vpl$) to display the CP
violating effects under CPT theorem \cite{CG-NPB,Aliev}. Moreover,
by considering the decaying chain $K^*\to K \pi$, we can investigate
various asymmetric operators by studying the angular distributions
of the $\pi$ meson \cite{CG-NPB}. In the following analysis, we will
concentrate on BRs and FBAs.

The paper is organized as follows. In Sec.~II, we first present the
mechanism to generate the process $b\to s \ell^{+} \ell^{-}$ in
the scenario of split SUSY. We then derive the differential
decay rates and FBAs with the new interactions. In Sec.~III,
by combining the results of $Br(B_{s}\to \mu^{+} \mu^{-})$ and the
solar neutrino mass, we estimate the decay rates and FBAs of $B\to
K^{(*)}\ell^+ \ell^-$.
 Finally, we
give our conclusions in Sec.~IV.

\section{Differential decay rates for $B\to K^{(*)} \ell^{+} \ell^{-}$}

\subsection{ $b\to s \ell^{+} \ell^{-}$ in split SUSY without R-parity}

In ordinary split supersymmetric models, due to the suppression of the
high SUSY breaking scale,  one expects that there will be no
interesting contributions to low energy physics. Therefore, we
extent our consideration to the framework of split SUSY with R-parity violation,
in which the conservations of lepton and baryon numbers are
broken. For simplicity, in this paper we only consider the
lepton number violating effects. The bilinear and trilinear
terms for the lepton number violation in the superpotential are
written as \cite {KO,CP}
 \be
 W=\mu H_1 H_2+\epsilon_i \mu L_i H_2+\lambda^{\prime}_{ijk} L_i Q_j
 D^{c}_{k} + \lambda_{ijk} L_{i} L_{j} E^{c}_{k}, \label{eq:w}
  \en
and the relevant scalar potential is given by
   \be
   V=BH_1H_2+B_{i}L_{i}H_{2}+m^{2}_{L_i H_1}L_i H^{\dagger}_{1}
   +h.c.\, . \label{eq:v}
   \en
Note that, we have used the same notations for superfields and
ordinary fields. The soft parameters $B$, $B_{i}$ and $m^{2}_{L_i
H_1}$ could be the same order as the SUSY breaking scale. It has
been shown  that to solve the atmospheric neutrino mass
$\sqrt{\Delta m^2_{atom}} \sim 0.05$ eV, the bilinear related
parameters of $\xi_{i}=m^2_{L_i H_1}/m^{2}_{L_i}+B_i/m^{2}_{L_i}
\tan\beta -\epsilon_i$ are limited to $10^{-6}/\cos\beta$ at tree
level \cite{CP}, where $\tan\beta=\langle H^{0}_{2}\rangle / \langle
H^0_1 \rangle$. In order to reach the solar neutrino mass scale of
$\sqrt{\Delta m^2_{sol}} \sim 9$ meV by the same bilinear couplings,
one has to go to one-loop level. However, in split SUSY the results
are suppressed by $ 1/m^{2}_{L_i}$. It is found that if the
trilinear R-parity violating couplings $\lambda^{\prime}_{i23}$ and
$\lambda^{\prime}_{i32}$ are of order one, the problem could be
solved by one-loop corrections \cite{CP}.

\begin{figure}[hpbt]
\includegraphics*[width=2.5in]{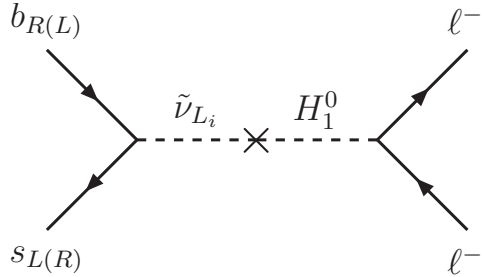} \caption{Tree contribution
to $b\to s \ell^+\ell^-$ with the cross representing the mixings
between sleptons and Higgs.}
 \label{tree1}
\end{figure}

Based on the above discussions, if we regard that
$\lambda^{\prime}_{i23,i32}$ as well as the ratios of the bilinear
couplings and $m_S^2$, i.e., $m^{2}_{L_i H_1}/m^2_{S}$ and
$B_i/m^2_{S}$, are order of unity, we find that
 $b \to s \ell^{+} \ell^{-}$ can occur at tree level shown
in Fig. \ref{tree1}, which may not be suppressed.
Since in the split SUSY approach, except the SM-like Higgs denoted
by $h^0$ is light, all scalars are extremely heavy, we
may simplify the calculations by using $-h^0 \sin\alpha\, (h^0 \cos
\alpha)$ instead of the Higgs $H^{0}_{1} (H^0_{2})$, where the
angle $\alpha$ describes the mixing of two neutral Higgses
\cite{Higgs}. From Eqs. (\ref{eq:w}) and (\ref{eq:v}), the
new effective interactions for $b\to s \ell^{+} \ell^{-}$ can be
described by
   \be
    {\cal H}_{\rm new}=
       \frac{1}{m^{2}_{h}}\left(\frac{gm_{\ell}}{2m_{W}} \frac{\sin\alpha }{\cos\beta}\right)
       \frac{m^{2}_{L_{i}
       H_1} }{m^{2}_{\tilde{\nu}_i}}
               \left(\lambda^{\prime*}_{i23} \bar{s} P_{R}+  \lambda^{\prime}_{i32} \bar{s} P_L b \right)
               \, \bar{\ell}\, \ell\, , \label{eq:bsll}
   \en
where $m_h$, $m_W$ and $m_{\ell}$ stand for the masses of Higgs,
W-boson and lepton, respectively, and $m^2_{L_iH_1}$ are from the
mixings between sleptons and Higgs.
Our purpose of this study is to examine
the influence of Eq. (\ref{eq:bsll}) on $B\to K^{(*)} \ell^{+}
\ell^{-}$ decays.

\subsection{Effective Hamiltonian and form factors for $B\rightarrow K^{(*)}$ transitions}
In order to include the new interactions for $b\to s \ell^{+}
\ell^{-}$, we write the effective Hamiltonian with interactions of
scalar and pseudoscalar to leptons as
  \be
 {\cal H}_{\rm eff}= \frac{G_F\alpha_{em} \lambda_t}{\sqrt{2}
 \pi}\left[ H_{1\mu} L^{\mu} +H_{2\mu}L^{5\mu} + S_R \bar\ell \ell +S_L \bar\ell \ell \right]
 \label{heff}
  \en
  with
  \begin{eqnarray}
  H_{1\mu } &=&C^{\rm eff}_{9}(\mu )\bar{s}\gamma _{\mu }P_{L}b\ -\frac{2m_{b}}{%
 q^{2}}C_{7}(\mu )\bar{s}i\sigma _{\mu \nu }q^{\nu }P_{R}b \,,
\nonumber \\
 H_{2\mu } &=&C_{10}\bar{s}\gamma _{\mu }P_{L}b \,,
 \nonumber\\
 L^{\mu } &=&\bar{\ell}\gamma ^{\mu }\ell\,, \ \ \ L^{5\mu } =\bar{\ell}\gamma ^{\mu }\gamma
 _{5}\ell\,,
 \nonumber\\
 S_R&=& C_R \bar{s} P_R b\, , \ \ \ S_L=C_L \bar{s} P_L b\, ,
 \label{heffc}
  \end{eqnarray}
  where $\alpha_{em}$ is fine structure constant, $\lambda_t=V_{tb}V_{ts}^*$,  $C^{\rm eff}_{9}$ and $C_{7,10}$
  are the Wilson coefficients (WCs) with their explicit
  expressions given in
 Ref.~\cite{BBL} for the SM,
  $C_{L(R)}$ are from the new interactions of split SUSY,
 $m_b$ is the current b-quark mass, $q$ is the momentum transfer and $P_{L(R)}=(1\mp \gamma_5)/2$. Note that $C^{\rm eff}_9$ has
 included the long-distance
 effects of $c\bar{c}$ bound states \cite{CGPRD66}.

To obtain the transition elements of $B\rightarrow H\left( H=K,\
K^{*}\right) $ with various weak vertices, we parametrize them in
terms of the relevant form factors as follows:
\begin{eqnarray}
\langle K(p_{2})| V_{\mu }| \bar{B} (p_{1})\rangle &=&
f_{+}(q^2)\Big\{P_{\mu}-\frac{P\cdot q }{q^2}q_{\mu} \Big\}
+\frac{P\cdot q}{q^2}f_{0}(q^2)\,q_{\mu},  \nonumber \\
\langle K(p_{2} )| T_{\mu }q^{\nu}| \bar{B} (p_{1})\rangle &=&
{f_{T}(q^2)\over m_{B}+m_{K}}\Big\{P\cdot q\,
q_{\mu}-q^{2}P_{\mu}\Big\},
\nonumber  \\
\langle K^{*}(p_{2},\epsilon )| V_{\mu }| \bar{B}%
(p_{1})\rangle &=&i\frac{V(q^{2})}{m_{B}+m_{K^{*}}}\varepsilon
_{\mu
\alpha \beta \rho }\epsilon ^{*\alpha }P^{\beta }q^{\rho },  \nonumber \\
\langle K^{*}(p_{2},\epsilon )| A_{\mu }| \bar{B}
(p_{1})\rangle &=&2m_{K^{*}}A_{0}(q^{2})\frac{\epsilon ^{*}\cdot q}{%
q^{2}}q_{\mu }+( m_{B}+m_{K^{*}}) A_{1}(q^{2})\Big( \epsilon
_{\mu }^{*}-\frac{\epsilon ^{*}\cdot q}{q^{2}}q_{\mu }\Big)  \nonumber \\
&&-A_{2}(q^{2})\frac{\epsilon ^{*}\cdot q}{m_{B}+m_{K^{*}}}\Big( P_{\mu }-%
\frac{P\cdot q}{q^{2}}q_{\mu }\Big) ,  \nonumber \\
\langle K^{*}(p_{2},\epsilon )| T_{\mu \nu }q^{\nu }| \bar{B}
(p_{1})\rangle &=&-iT_{1}(q^{2})\varepsilon _{\mu \alpha \beta
\rho
}\epsilon ^{*\alpha }P^{\beta }q^{\rho },  \nonumber \\
\langle K^{*}(p_{2},\epsilon )| T_{\mu \nu }^{5}q^{\nu }|
\bar{B}(p_{1})\rangle &=&T_{2}(q^{2})\Big( \epsilon _{\mu
}^{*}P\cdot q-\epsilon ^{*}\cdot qP_{\mu }\Big)
+T_{3}(q^{2})\epsilon ^{*}\cdot q\Big( q_{\mu
}-\frac{q^{2}}{P\cdot q}P_{\mu }\Big)\, , \label{ffv}
\end{eqnarray}
where $(V_{\mu },A_{\mu },T_{\mu \nu },T_{\mu \nu }^{5})=\bar{s}(\gamma _{\mu }, \gamma_{\mu }\gamma _{5}, i\sigma _{\mu \nu }, i\sigma _{\mu \nu }\gamma _{5})b$,
$m_{B,K,K^*}$ are the meson masses of $B$, $K$ and
$K^*$,  $P=p_{B}+p_{K^{(*)}}$,
$q=p_{B}-p_{K^{(*)}}$ and $P \cdot q=m^{2}_{B}-m^{2}_{K^{(*)}}$,
respectively.
By equation of motion, we can have the transition form factors for
scalar and pseudoscalar interactions as
  \be
   \langle K| \bar{s} \; b| \bar{B} \rangle & = & \frac{P\cdot q}{(m_s-m_b)}\, f_0,
   \non \\
   \langle K^*| \bar{s} \gamma_5 b| \bar{B} \rangle & = & - \frac{2m_{K^*}}{m_b + m_s}
    \epsilon^*\cdot q\, A_0\,,
    \label{sff}
  \en
 with $m_{s}$ being the strange-quark mass.

\subsection{Angular Distributions and Forward-Backward Asymmetries}

 From the definitions of form factors in Eqs. (\ref{ffv}) and (\ref{sff}),
the transition amplitudes associated with the interactions in
Eq.~(\ref{heff}) for $B\rightarrow K^{(*)}\ell^{+}\ell^{-}$  can
be written as
 \be
       {\cal M}_{K}&=&\frac{G_{F}\alpha_{em} \lambda _{t}}{\sqrt{2}\pi }
       \left[ m_{97} \bar{\ell} \not{p}_K \ell + m_{10} \bar{\ell} \not{p}_K \gamma_5 \ell
       +m_5 \bar{\ell} \gamma_5 \ell +m_0 \bar{\ell} \ell
       \right]\label{tampk}
 \en
 with
 \be
  m_{97}&=& C^{\rm eff}_9 f_+ +\frac{2m_b}{m_B+m_K}C_7
  f_T \,,\non \\
  m_{10}&=& C_{10} f_+\,, \non \\
  m_5&=& m_{\ell}C_{10}\left( f_+ +\frac{P\cdot q}{q^2}(f_0 -
  f_+)\right)\,,\non\\
  m_0 &=& \frac{C_R+C_L}{2}\frac{P\cdot q}{m_s-m_b} f_0\, ,
  \label{TMK}
 \en
and

\begin{equation}
{\cal M}_{K^{*}}=\frac{G_{F}\alpha \lambda _{t}}{\sqrt{2}%
\pi }\left\{ {\cal M}_{1\mu }\bar{\ell}\gamma^{\mu} \ell+{\cal
M}_{2\mu } \bar{\ell}\gamma^{\mu}\gamma_5 \ell + \tm_5
\frac{\epsilon^* \cdot q}{q^2} \bar{\ell} \gamma_5 \ell + \tm_0
\epsilon^*\cdot q\bar{\ell} \ell\right\} \label{ampk*}
\end{equation}
where
\begin{eqnarray}
{\cal M}_{1\mu } &=& i\tm^{1}_{97}\varepsilon _{\mu \nu \alpha
\beta }\epsilon ^{*\nu }p^{\alpha }_{K} q^{\beta }-\tm^2_{97}
\epsilon _{\mu }^{*}+\tm^{3}_{97}\epsilon ^{*}\cdot q p_{K\mu },
\non \\
{\cal M}_{2\mu } &=& i\tm^{1}_{10}\varepsilon _{\mu \nu \alpha
\beta }\epsilon ^{*\nu }p^{\alpha }_{K} q^{\beta }-\tm^2_{10}
\epsilon _{\mu }^{*}+\tm^{3}_{10}\epsilon ^{*}\cdot q p_{K\mu },
 \label{ampk1}
\end{eqnarray}
with
      \be
      \tm^{1}_{97}&=& \frac{V}{m_B+m_{K^*}} C^{\rm eff}_9
      +\frac{2m_b}{q^2}C_7 T_1 \, ,\non\\
       \tm^{2}_{97}&=& \frac{1}{2}(m_B+m_{K^*}) C^{\rm eff}_9 A_1
      +\frac{1}{2} \frac{2m_b}{q^2}P\cdot q C_7 T_2 \, ,\non\\
      \tm^{3}_{97}&=& \frac{A_2}{m_B+m_{K^*}} C^{\rm eff}_9
      +\frac{2m_b}{q^2}C_7 \left(T_2+\frac{q^2}{P\cdot q}T_3\right) \, ,\non\\
      \tm^i_{10}&=&\tm^{i}_{97}|_{C_7\to 0,C_9^{eff}\to C_{10}}\,,\non\\
      \tm_5&=& m_{\ell} C_{10} \left[-2m_{K^*} A_0 +(m_B+m_{K^*}) A_1 -\frac{q^2+P\cdot q}{m_B+m_{K^*}}A_2
      \right]\,, \non \\
      \tm_0&=& (C_L-C_R)\frac{m_{K^*}}{m_b+m_s}A_0\,. \label{TMKs}
      \en

To get the decay rate distributions in terms
of the dilepton invariant mass $q^2$ and the lepton
polar angle $\theta$, we use the $q^2$ rest frame in which
 $p_{\ell}=(E_{\ell}, |\vpl|\sin\theta, 0,
|\vpl|\cos\theta)$, $p_{H}=(E_{H},0,0,|\vec{p}_{H}|\cos\theta)$
with $E_{\ell}=\sqrt{q^2}/2$,
$|\vpl|=\sqrt{E^{2}_{\ell}-m^{2}_{\ell}}$,
$E_{H}=(m^{2}_{B}-q^2-m^2_{H})/(2\sqrt{q^2})$ and
$|\vec{p}_{H}|=\sqrt{E^2_{H}-m^{2}_{H}}$. By squaring the
transition amplitude in Eq.~(\ref{tampk}) and including the
three-body phase space factor, the differential decay rates as
functions of $q^2$ and $\theta$ for
$B\to K \ell^+ \ell^-$ are given by
 \be
\frac{d\Gamma _{K} }{dq^2
d\cos\theta}&=&\frac{G_{F}^{2}\alpha^{2}_{em}| \lambda _{t}|
^{2}}{ 2^{8}m^{2}_B \pi
^{5}}\sqrt{1-\frac{(2m_{\ell})^2}{q^2}}\tilde{p}_{K}\non\\
& \times & \left[ |\vpk|^2 \left(q^2-4|\vpl|^2 \cos^2\theta
\right)\left( |m_{97}|^2+|m_{10}|^2\right)+4m^{2}_{K}m^2_{\ell}
|m_{10}|^2
\right.\non\\
&+&  q^2\left( |m_5|^2+|m_{0}|^2\left(
1-\frac{(2m_{\ell})^2}{q^2}\right) \right)+ 4 m_{\ell} q\cdot
p_{K} Re(m_{10} m^*_5 )\non \\
&+& \left. 8m_{\ell} |\vpk| |\vpl| Re(m_{97}m^*_{0}) \cos\theta
\right]. \label{difk}
 \en
 For $B\rightarrow K^* \ell^{+} \ell^{-}$ decays, by summing up the polarizations of $K^*$ with the identity
 $\sum \epsilon^{*}_{\mu}(p) \epsilon_{\nu}(p) = (-g_{\mu\nu}+p_{\mu}p_{\nu}/p^2)$,
 from Eq. (\ref{ampk*}) the
differential decay rates
are found to be
 \begin{eqnarray}
\frac{d\Gamma_{K^*} }{ dq^{2}d\cos \theta} &=&\frac{
G_{F}^{2}\alpha^{2}_{em}|\lambda _{t}| ^{2} }{2^{8}m_{B}^{2} \pi
^{5}} \sqrt{1-\frac{(2m_{\ell})^2}{q^2}}\tilde{p}_{K^*}   \left\{
q^2 |\vpks|^2 \left(q^2+4|\vpl|^2\cos^2\theta
 \right) \left(|\tm^1_{97}|^2+|\tm^1_{10}|^2 \right) \right. \non
 \\
 &+&  4m^2_{\ell} q^2 |\vpks|^2 \left(|\tm^1_{97}|^2 - |\tm^1_{10}|^2 \right)
 + \frac{ |\vpks|^2}{m^2_{K^*}}  \left( q^2-4 |\vpl|^2 \cos^2\theta \right)
 \left(|\tm^2_{97}|^2+|\tm^2_{10}|^2 \right) \non \\
 &+&  2q^2 \left[\left( 1+\frac{2m^2_{\ell}}{q^2}\right)|\tm^2_{97}|^2
 + \left( 1-\frac{(2m_{\ell})^2}{q^2}\right)|\tm^2_{10}|^2
 \right] \non \\
 &+&  \frac{q^2}{m^2_{K^*}} |\vpks|^4 \left( q^2 -4|\vpl|^2 \cos^2\theta
 \right) \left( |\tm^3_{97}|^2 + |\tm^3_{10}|^2 \right)+4
 m^2_{\ell} q^2 |\vpks|^2 |\tm^3_{10}|^2 \non \\
 &-&  2\frac{q\cdot p_{K^*}}{m^2_{K^*}} |\vpks|^2 \left( q^2 -4 |\vpl|^2 \cos^2\theta
 \right) \left( Re(m^2_{97} m^{3*}_{97}) + Re(m^2_{10}
 m^{3*}_{10})\right) \non \\
 &+&   \frac{|\vpks|^2}{m^2_{K^*}} \left[ |\tm_5|^2 +
  q^4 \left(
 1-\frac{(2m_{\ell})^2}{q^2}\right)|\tm_{0}|^2
 -4 m_{\ell}  \left( Re\left(\tm^2_{10}-\tm^3_{10}q\cdot p_{K^*}\right)\tm^*_5
\right) \right]  \non \\
  &-& 8 |\vpks||\vpl|q^2 \left[Re(\tm^1_{97}m^{2*}_{10})+ Re( \tm^2_{97}m^{1*}_{10})
  \right] \cos\theta \non \\
  &-& \left. 8 m_{\ell} \frac{|\vpks||\vpl|}{m^2_{K^*}} \left[
  q\cdot P_{K^*} Re(\tm^2_{97} \tm^*_{0}) -q^2 |\vpks|^2 Re(\tm^3_{97} \tm^*_{0})
  \right] \cos\theta
  \right\}\label{difks}.
\end{eqnarray}
Note that, to have general formulas,
we have kept the lepton mass effects in Eqs. (\ref{difk}) and (\ref{difks}). The
$\tilde{p}_{H}$ in both differential decay rates are the spatial
momentum of the $H$ meson in the $B$-meson rest frame, defined as $
\tilde{p}_{H}=\sqrt{ E^{\prime 2}- m^{2}_{H} }$ with $E^{\prime
}=(m^{2}_{B}+m^{2}_{H}-q^{2})/(2m_{B})$.

By the angular distributions, we can define interesting physical
observables, such as the FBAs, given by
  \be
  {\cal A}^{FB}_{H}={ -\int_{0}^{\pi/2}
  \frac{d\Gamma_{H}}{dq^2 d\cos\theta} d\cos\theta
  + \int_{\pi/2}^{\pi} \frac{d\Gamma_{H}}{dq^2 d\cos\theta} d\cos\theta
  \over \int_{0}^{\pi/2}
  \frac{d\Gamma_{H}}{dq^2 d\cos\theta} d\cos\theta
  + \int_{\pi/2}^{\pi} \frac{d\Gamma_{H}}{dq^2 d\cos\theta}
  d\cos\theta}. \label{eq:fba}
  \en
It was pointed out that the FBAs of $B\to K^* \ell^{+} \ell^{-}$
are associated with the difference between transverse
polarizations of $K^*$ \cite{CG-NPB}, denoted simply by $Re({\cal M}^{+}_{1}
{\cal M}^{+}_{2})-Re({\cal M}^{-}_{1} {\cal M}^{-}_{2})$ with
 \be
   {\cal M}^{\pm}_{1}=\pm |\vpks|\sqrt{q^2} \tm^{1}_{97} +
   \tm^{2}_{97}\, , \non \\
    {\cal M}^{\pm}_{2}=\pm |\vpks|\sqrt{q^2} \tm^{1}_{10} +
   \tm^{2}_{10}.
 \en
Clearly, the resultant of $Re(\tm^1_{97}m^{2*}_{10})+ Re(
\tm^2_{97}m^{1*}_{10})$ is expected in the SM.
For $B\to K \ell^{+} \ell^{-}$,
since there are no transverse degrees of freedom, we can infer that there are
no FBAs in the SM. Therefore,
nonzero FBAs for $B\to K \ell^{+} \ell^{-}$ can be
strong evidences of new physics and they
have to be from the longitudinal parts. Indeed, from
Eq.~(\ref{difk}), it is understood that  FBA terms are
related to $Re(m_{97}m^{*}_{0})$, in which $m^{*}_{0}$ is induced from
the new scalar interactions. In addition, since the associated
FBAs are the interference effects of vector and scalar currents, the
mass factor $m_{\ell}$ also appears to get the correct chirality.

\section{Numerical calculations}

\subsection{Constraints from  the decay of
$B_s \to \mu^{+} \mu^{-}$ and the mass of solar neutrino}
\label{sub:limit}

 From Eq. (\ref{eq:bsll}), we first consider the $B_{s}\to \ell^{+}
\ell^{-}$ decay, which
 contains few hadronic uncertainties. The corresponding
decay amplitude is given by
    \be
       A=\langle \ell^{+} \ell^{-} | H_{\rm new} |\bar{B}_s\rangle =
       \frac{-i}{2m^{2}_{h}}\left(\frac{gm_{\ell}}{2m_{W}} \frac{\sin\alpha }{\cos\beta}\right)
       \frac{f_{B_s} m^{2}_{B_s} }{(m_b+m_s)}\frac{m^{2}_{L_{i}
       H_1} (\lambda^{\prime*}_{i23}-
        \lambda^{\prime}_{i32})}{m^{2}_{\tilde{\nu}_i}}\, \bar{\ell}\,
        \ell,
              \label{eq:ampbll}
    \en
 where we have used the identity $\langle0 | \bar{s} \gamma_5
b|\bar{B}_s\rangle\approx -i\, f_{B_s} m^2_{B_s}/(m_b+m_s)$ with
$f_{B_s}$ being the decay constant. Since the trilinear couplings
in sleptons and quarks involve two possible chiralities, there is
a cancelation in Eq.~(\ref{eq:ampbll}). Note that if
$\lambda^{\prime*}_{i23}=\lambda^{\prime}_{i32}$, our mechanism
vanishes automatically. By including the phase space factor, the
decay rate is given by
       \be
         \Gamma&=&
         \frac{m_{B_s}}{16\pi }
         \frac{G_{F} m^{2}_{\ell}}{\sqrt{2}}  \left( \frac{f_{B_s}m_{B_s}
         }{m^{2}_{h}} \frac{\sin\alpha}{\cos\beta}
        |{\cal N}_i| \right)^2
          \left[ 1- \left( \frac{2m_{\ell}}{m_{B_s}} \right)^2
         \right]^{3\over 2}
       \label{Rate}
       \en
with ${\cal N}_i=m^{2}_{L_{i}H_1}(\lambda^{\prime*}_{i23}-
\lambda^{\prime}_{i32})/ m^{2}_{\tilde{\nu}_i}$. In the SM, it is
known that $B_s\to \ell^{+} \ell^{-}$  arises from the
electroweak penguin and box diagrams. The decay
BR of  $B_s\to \mu^{+} \mu^{-}$ is found to be $(3.8 \pm
1.0)\times 10^{-9}$ \cite{BurasPLB} which is much less than the
current experimental upper limit of $5.0\times 10^{-7}$ \cite{D0}.
In order
to further limit the values of unknown parameters, we include the solar
neutrino mass which is presented by \cite{CP}
  \be
   M^{\nu}_{ij}\sim \frac{3}{8\pi^2}\lambda^{\prime}_{i23}
   \lambda^{\prime}_{j32}\frac{m_b m_s}{m_S}.
  \en
  To preserve the solar neutrino mass to be $\sim 9$ meV,
  we choose $\lambda^{\prime}_{i23}=0.9$,
  $\lambda^{\prime}_{i32}=0.5$, $m_b=4.5$ GeV,  $m_s=0.13$ GeV, and
  $m_{S}=10^{9}$ GeV. Since the remaining free parameters are $m_{L_i
  H_1}/m_{\tilde{\nu}_i}$ and $m_{h}$,
  to illustrate our numerical results, we set $m_{L_i
  H_1}/m_{\tilde{\nu}_i}=0.2$ and $m_{h}=150$ GeV and take $m_{B_s}=5.37$ GeV,
  $f_{B_s}=0.23$ GeV, $\tau_{B_s}=1.46\times 10^{-12}\, s$ and
  $\alpha=\pi/2+\beta$. As a result,
   we get $Br(B_{s}\to \mu^{+} \mu^{-})=1.66\times
  10^{-8}$ which is one order of magnitude larger than that
  predicted in the SM. It is clear that a heavier Higgs or a smaller $m_{L_i
  H_1}/m_{\tilde{\nu}_i}$ will make the contribution be
  smaller. In the following numerical calculations, we
  will adopt the above values of parameters to estimate BRs and
  FBAs in $B\to K^{(*)} \ell^{+} \ell^{-}$ decays.

\subsection{Decay rate distributions and branching ratios}
\begin{figure}[hpbt]
\includegraphics*[width=4.in]{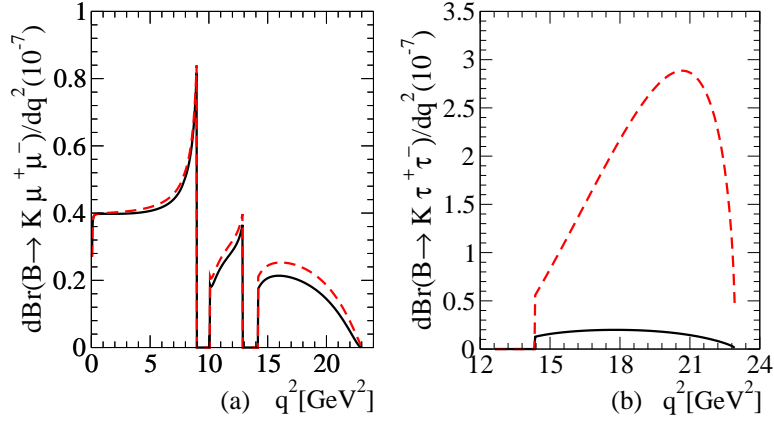} \caption{Differential BRs (in
units of $10^{-7}$) for (a) $B\to K^{0} \mu^{+} \mu^{-}$ and
(b) $B \to K^{0} \tau^{+} \tau^{-}$, where the dashed and solid
 lines correspond to the results with and without new
physics, respectively.}
 \label{fig:brk}
\end{figure}

\begin{figure}[hpbt]
\includegraphics*[width=4.in]{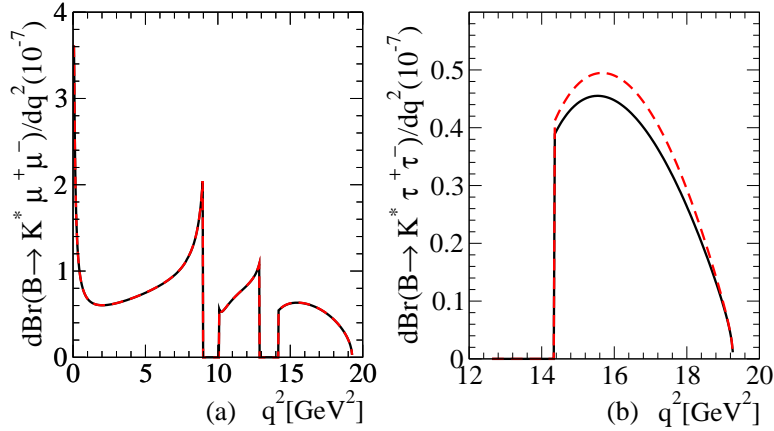} \caption{Same as Fig. \ref{fig:brk}
but for
(a) $B\to K^{*0} \mu^{+} \mu^{-}$ and
(b) $B \to K^{*0} \tau^{+} \tau^{-}$.}
 \label{fig:brks}
\end{figure}

By integrating the lepton polar angle, the differential
decay rate as a function of the dilepton invariant mass for $B\to K
\ell^+ \ell^-$ is given by
 \be
\frac{d\Gamma _{K} }{dq^2}&=&\frac{G_{F}^{2}\alpha^{2}_{em}|
\lambda _{t}| ^{2}}{ 2^{8}m^{2}_B \pi
^{5}}\sqrt{1-\frac{(2m_{\ell})^2}{q^2}}\tilde{p}_{K}\non\\
& \times & \left[ \frac{4}{3} |\vpk|^2 \left(q^2+2m^2_{\ell}
\right)\left( |m_{97}|^2+|m_{10}|^2\right)+ 8 m^{2}_{K}m^2_{\ell}
|m_{10}|^2
\right.\non\\
&+& \left.2 q^2\left( |m_5|^2+|m_{0}|^2\left(
1-\frac{(2m_{\ell})^2}{q^2}\right) \right)+ 8 m_{\ell} q\cdot
p_{K} Re(m_{10} m^*_5 ) \right]. \label{difkq2}
 \en
Similarly,
for $B\to K^* \ell^{+} \ell^{-}$, one has
\begin{eqnarray}
\frac{d\Gamma_{K^*} }{ dq^{2}} &=&\frac{
G_{F}^{2}\alpha^{2}_{em}|\lambda _{t}| ^{2} }{2^{8}m_{B}^{2} \pi
^{5}} \sqrt{1-\frac{(2m_{\ell})^2}{q^2}}\tilde{p}_{K^*} \left\{
\frac{8}{3}q^2 |\vpks|^2 \left(q^2 - m^2_{\ell}
 \right) \left(|\tm^1_{97}|^2+|\tm^1_{10}|^2 \right) \right. \non
 \\
 &+&  8m^2_{\ell} q^2 |\vpks|^2 \left(|\tm^1_{97}|^2 - |\tm^1_{10}|^2 \right)
 +\frac{4}{3} \frac{ |\vpks|^2}{m^2_{K^*}}  \left( q^2+2m^2_{\ell} \right)
 \left(|\tm^2_{97}|^2+|\tm^2_{10}|^2 \right) \non \\
 &+&  4q^2 \left[\left( 1+\frac{2m^2_{\ell}}{q^2}\right)|\tm^2_{97}|^2
 + \left( 1-\frac{(2m_{\ell})^2}{q^2}\right)|\tm^2_{10}|^2
 \right] \non \\
 &+&  \frac{4}{3}\frac{q^2}{m^2_{K^*}} |\vpks|^4 \left( q^2 +2 m^2_{\ell}  \right)
  \left( |\tm^3_{97}|^2 + |\tm^3_{10}|^2 \right)+ 8
 m^2_{\ell} q^2 |\vpks|^2 |\tm^3_{10}|^2 \non \\
 &-&  \frac{8}{3}\frac{q\cdot p_{K^*}}{m^2_{K^*}} |\vpks|^2 \left( q^2 +2m^2_{\ell}  \right)
 \left( Re(m^2_{97} m^{3*}_{97} + Re(m^2_{10}
 m^{3*}_{10})\right) +2\frac{|\vpks|^2}{m^2_{K^*}} \non \\
 &\times& \left.  \left[ |\tm_5|^2 +
  q^4 \left(
 1-\frac{(2m_{\ell})^2}{q^2}\right)|\tm_{0}|^2
 -4 m_{\ell}  \left( Re\left(\tm^2_{10}-q\cdot p_{K^*}\tm^3_{10}\right)\tm^*_5
\right) \right]  \right\} \label{difksq2}.
\end{eqnarray}
To calculate the numerical values, we use $m_{B}=5.28$ GeV,
$m_K=0.5$ GeV, $m_{K^*}=0.89$ GeV,
$\alpha=1/129$, $V_{ts}=0.40\pm 0.003$, $\tau_{B^0}=1.536\times
10^{-12}$ s and $\tau_{B^+}=1.671\times 10^{-12}$ s. For the
form factors in $\langle K^{(*)}| \Gamma| \bar{B}\rangle$, we
quote the updated results of the light-cone sum rules (LCSR)
\cite{Ball}. The explicit fitting forms are summarized as
   \be
    f_{+}(q^2)&=&\frac{0.1903}{1-q^2/29.3}+\frac{0.1478}{(1-q^2/29.3)^2},
    \ \ \ f_{0}(q^2)= \frac{0.3338}{1-q^2/38.98},\non \\
    f_{T}(q^2)&=&
    \frac{0.1851}{1-q^2/29.3}+\frac{0.1905}{(1-q^2/29.3)^2}, \non
    \\
    V(q^2)&=&\frac{0.923}{1-q^2/5.32^2}-\frac{0.511}{1-q^2/49.4}, \
    \ \
    A_{0}(q^2)=\frac{1.364}{1-q^2/5.282^2}-\frac{0.99}{1-q^2/36.78},
    \non \\
    A_{1}(q^2)&=& \frac{0.29}{1-q^2/40.38}, \ \ \
    A_{2}(q^2)=-\frac{0.084}{1-q^2/52.}+\frac{0.342}{(1-q^2/52.)^2},\non\\
    T_{1}(q^2)&=&\frac{0.823}{1-q^2/5.32^2}-\frac{0.491}{1-q^2/46.31},
    \ \ \ T_{2}(q^2)=\frac{0.333}{1-q^2/41.41}, \non \\
    T_{3}(q^2)&=&-\frac{0.036}{1-q^2/48.1}+\frac{0.368}{(1-q^2/48.1)^2}\, .
   \en
In order to exclude the backgrounds from $B\to J/\Psi
(\Psi^{\prime}) K^{(*)}$, we follow the BELLE's veto windows
\cite{Bellebsll}, defined
as:
    \be
       {\rm I_V}:&& \ \ \ -0.20{\rm GeV}<M_{e^+e^-}-m_{V}< 0.07 {\rm
       GeV}, \non \\
       {\rm II_V}:&& \ \ \ -0.10{\rm GeV}<M_{\mu^+ \mu^-}-m_{V}< 0.08 {\rm
       GeV}  \label{veto1}
           \en
  for $B\to K \ell^{+} \ell^{-}(\ell=e,\,\mu)$  with $V=\Psi,\, \Psi^{\prime}$
  and for $B\to K^* \ell^{+} \ell^{-}$ ($\ell=e,\mu$) with $V=\Psi^{\prime}$, and
     \be
       {\rm III_{\Psi}}:&& \ \ \ -0.25{\rm GeV}<M_{e^+e^-}-m_{J/\Psi}< 0.07 {\rm
       GeV}, \non \\
       {\rm IV_{\Psi}}: && \ \ \ -0.15{\rm GeV}<M_{\mu^+ \mu^-}-m_{J/\Psi}< 0.08 {\rm
       GeV} \label{veto2}
           \en
  for $B\to K^* \ell^{+} \ell^{-} (\ell=e,\, \mu)$ decays. For
  $B\to K^{(*)} \tau^{+} \tau^{-}$,
   we use
   \be
   V_{\Psi^{\prime}}:&&\, M_{\tau^{+} \tau^{-}}-m_{\Psi^{\prime}} <
  0.08\,.
  \label{veto3}
  \en
 \begin{table}[hptb]
\caption{\label{table:br} The average BRs (in units of $10^{-7}$) of
$B\to K^{(*)} \ell^{+} \ell^{-}$ for charged and neutral $B$ decays
in the SM with the veto windows defined in Eqs.~(\ref{veto1}),
(\ref{veto2}) and (\ref{veto3}).}
\begin{ruledtabular}
\begin{tabular}{cccc}
 Mode  & $B\to K e^+ e^-$ &$B\to K \mu^+ \mu^- $& $B\to K \tau^+ \tau^-$  \\ \hline 
  BR & $5.59 $ & $6.18$ & $1.46$  \\ \hline 
  Mode & $B\to K^{*} e^+ e^-$ &$B\to K^{*} \mu^+ \mu^- $& $B\to K^{*} \tau^+ \tau^-$
  \\ \hline 
  BR & $14.79$ & $12.54$ & $1.75$
\end{tabular}
\end{ruledtabular}
\end{table}
Hence, the predicted values in the SM for various lepton modes are
presented in Table~\ref{table:br}. Note that, the values in the
table are obtained by taking the average lifetime in charged and
neutral $B$ mesons. Furthermore, according to the results of
Table~\ref{table:br}, the average BRs for the $\ell=e$ and $\mu$
modes in the SM are given by
  \be
    Br(B\to K \ell^{+} \ell^{-})&=&5.89\times 10^{-7} \,,\non \\
    Br(B\to K^* \ell^{+} \ell^{-})&=&13.67\times 10^{-7}\,.
  \en
In the
  split SUSY
R-parity violating model,
 we obtain
    \be
    Br(B\to K \ell^{+} \ell^{-})&=&6.14\times 10^{-7}\, , \non \\
    Br(B\to K \tau^{+} \tau^{-})&=&17.95\times 10^{-7}\, ,\non\\
    Br(B\to K^* \ell^{+} \ell^{-})&=&13.87\times 10^{-7}\, , \non \\
    Br(B\to K^* \tau^{+} \tau^{-})&=&1.91\times 10^{-7}.
  \en
    It is interesting to see that the decay BR of $B\to K \tau^{+} \tau^{-}$ gets  one
  order of magnitude enhancement, while that of $B\to K^{*} \tau^{+} \tau^{-}$
  changes only a little.
  Hence, the new scalar interactions could largely enhance the BR
  of $B\to K \tau^{+} \tau^{-}$, but not that of the $K^*$ mode.
 In Figs. \ref{fig:brk} and
  \ref{fig:brks}, we show the differential decay BRs of
  $B\to K^0\mu^+\mu^-$ and $K^0\tau^+\tau^-$,
  where the dashed and solid lines stand
  for the predictions with and without new physics, respectively.
  Since the scalar interactions are associated with the lepton mass,
   as displayed by the figures, the influences on light lepton modes are small.

\subsection{Forward-backward asymmetries}

It is known that FBAs of $b\to s \ell^{+} \ell^{-}$ decays could
be sensitive to new physics \cite{Ali,CG-NPB,CGPRD66,FBA}.
 From Eqs.~(\ref{difk}), (\ref{difks}) and
(\ref{eq:fba}),
the FBAs for $B\to K\ell^+\ell^-$ and $K^*\ell^+\ell^-$ are given by
  \be
{\cal A}_{K}(q^2) &=&
\frac{1}{d\Gamma_{K}/dq^2}\frac{G_{F}^{2}\alpha^{2}_{em}| \lambda
_{t}| ^{2}}{ 2^{8}m^{2}_B \pi
^{5}}\sqrt{1-\frac{(2m_{\ell})^2}{q^2}}\tilde{p}_{K} \Big[-
8m_{\ell} |\vpk| |\vpl| Re(m_{97}m^*_{0}) \Big]\, ,
\label{fbasyk}
 \\
   {\cal A}_{K^*}(q^2) &=&
\frac{1}{d\Gamma_{K^*}/dq^{2}}\frac{
G_{F}^{2}\alpha^{2}_{em}|\lambda _{t}| ^{2} }{2^{8}m_{B}^{2} \pi
^{5}}
\sqrt{1-\frac{(2m_{\ell})^2}{q^2}}\tilde{p}_{K^*} \non \\
&\times&  \Big\{
   8 |\vpks||\vpl|q^2 \left[Re(\tm^1_{97}m^{2*}_{10})+ Re( \tm^2_{97}m^{1*}_{10})
  \right]    \non \\
  &+& \left. 8 m_{\ell} \frac{|\vpks||\vpl|}{m^2_{K^*}} \left[
  q\cdot P_{K^*} Re(\tm^2_{97} \tm^*_{0}) -q^2 |\vpks|^2 Re(\tm^3_{97} \tm^*_{0})
  \right]
  \right\}\,,\label{fbasyks}
    \en
respectively.
As discussed before, the new contributions from scalar
interactions are suppressed due to the light lepton masses.
However, since ${\cal A}_{K}(q^2)$ vanishes in the SM, the search of the FBAs
in $B\to K \ell^{+} \ell^{-}$ with lots of accumulated $B$
samples is still important in future B factories, such as SuperB, BTeV and LHCB. 
\begin{figure}[hpbt]
\includegraphics*[width=4.in]{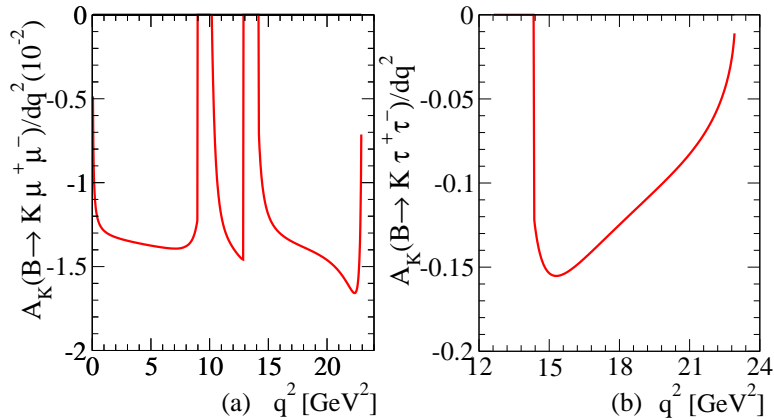} \caption{Forward-backward asymmetries (FBAs)
 for (a) $B\to K \mu^{+} \mu^{-}$ and
(b) $B\to K \tau^{+} \tau^{-}$.}
 \label{fig:fbak}
\end{figure}
For this reason,
we would display the FBAs of $B\to K \ell^{+} \ell^{-}$ ($\ell=\mu$ and $\tau$)
associated with the new interactions as functions of $q^2$ in
Fig.~\ref{fig:fbak}. Here, we have also used the criterions of
Eqs.~(\ref{veto1}), (\ref{veto2}) and  (\ref{veto3}) to
exclude the backgrounds.
 From the figure, we see clearly that ${\cal A}_{K}(B\to K \ell^{+}\ell^{-})$ can be as large as  $1$ and $10\%$ for $\ell=\mu$ and $\tau$, which
 require at least $2\times 10^{10}$ and $6\times 10^7$
B samples for experimental observations at $1\sigma$ level.
\begin{figure}[hpbt]
\includegraphics*[width=4.in]{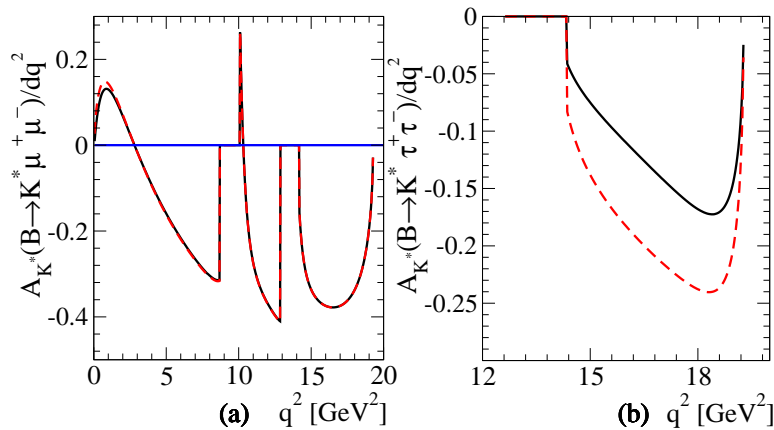} \caption{Forward-backward asymmetries (FBAs) for
(a) $B\to K^* \mu^{+} \mu^{-}$ and (b) $B\to K^* \tau^{+}
\tau^{-}$, where dashed and solid lines represent the results
with and without new physics, respectively.}
 \label{fig:fbaks}
\end{figure}
In Fig.~\ref{fig:fbaks}, we present the FBAs of $B\to K^* \ell^{+}
\ell^{-}$. From
Fig.~\ref{fig:fbaks}(a), we find that the new interactions have no
effect on ${\cal A}_{K^*}(B\to K^* \mu^{+} \mu^{-})$,
but  a large influence on ${\cal A}_{K^*}(B\to K^* \tau^{+}
\tau^{-})$.

\section{Conclusions}

We have studied the decays of $B_{s}\to \mu^{+} \mu^{-}$ and $B\to
K^{(*)} \ell^{+} \ell^{-}$ in split SUSY with R-parity violation.
With the new updated form factors calculated by the light-cone sum
rules, we have obtained the decay branching ratios within the
BELLE's veto windows to be $Br(B\to K^{(*)} \ell^{+}\ell^{-})=5.89\
(13.67)\times 10^{-7}$ for $\ell=e$ and $\mu$ in the SM. If we set
the  parameters of new interactions such that the BR of
$B_{s}\to \mu^{+} \mu^{-}$ is one order of magnitude larger than
that of SM, we find that the decay branching ratio of $B\to
K\tau^+\tau^-$ can be $1.8\times 10^{-6}$ which is about one order
of magnitude larger that in the SM, whereas those of $B\to
K\ell^+\ell^-$ ($\ell=e$ and $\mu$) and the $K^*$ modes are
insensitive to the new effects. For the FBAs, we have shown that
${\cal A}_K$ for $B\to K \tau^{+} \tau^{-}$ and $K\mu^{+}\mu^{-}$,
which are zero in the SM, can be over $10$ and $1\%$ in our new
physics model,  respectively. In addition, we have also demonstrated that
the new interactions can significantly change the spectrum of ${\cal
A}_{K^*}$ in $B\to K^* \tau^{+} \tau^{-}$. \\

{\bf Acknowledgments}\\

This work is supported in part by the National Science Council of
R.O.C. under Grant \#s: NSC-93-2112-M-006-010 and
NSC-93-2112-M-007-014.



\begin{thebibliography}{99}




\bibitem{unify1}S. Dimopoulos and H. Georgi, Nucl. Phys. B{\bf
193}, 150 (1981).

\bibitem{unify2} S. Dimopoulos, S. Raby and F. Wilczek, Phys. Rev.
D{\bf 24}, 1681 (1981).

\bibitem{dark}H. Goldberg, Phys. Rev. Lett. {\bf 50}, 1419
(1983).

\bibitem{AD} N. Arkani-Hamed and S. Dimopoulos, hep-th/0405159.

\bibitem{ADGR} N. Arkani-Hamed, S. Dimopoulos, G.F. Giudice and  A.
Romanino, Nucl. Phys. B{\bf 709},3 (2005).

\bibitem{GR}  G.F. Giudice and  A. Romanino, Nucl. Phys. B{\bf 699}, 65 (2004).

\bibitem{CG-hepph}C.H. Chen and C.Q. Geng, hep-ph/0501001.

\bibitem{Ali} A. Ali {\it et al.}, Phys. Rev. D{\bf 61}, 074024
(2000).

\bibitem{MNS}
D. Melikhov, N. Nikitin, and S. Simula, Phys. Rev. D{\bf 57}, 6814
(1998).

\bibitem{CG-PRD} C.H. Chen and C.Q. Geng, Phys. Rev. D{\bf 63}, 114025
(2001).

\bibitem{Bellebsll}
Belle Collaboration, K. Abe {\it et al.}, hep-ex/0410006.

\bibitem{BaBarbsll}
BABAR Collaboration, B.~Aubert {\it et al.},
 Phys. Rev. Lett. {\bf 91}, 221802 (2003).

\bibitem{lep}  F. Kruger and L. M. Sehgal, Phys. Lett. B{\bf 380}, 199
(1996);  J.L. Hewett, Phys. Rev. D{\bf 53}, 4964 (1996); C.Q. Geng
and C.P. Kao, Phys. Rev. D{\bf 54}, 5636 (1996);
C.Q. Geng and C.P. Kao, Phys. Rev. D{\bf 57}, 4479 (1998); T.M. Aliev {\it
et al.}, Phys. Rev. D{\bf 64}, 055007 (2001); W. Bensalam {\it et
al.},  Phys. Rev. D{\bf 67}, 034007 (2003); S.R. Choudhury {\it et
al.}, Phys. Rev. D{\bf 68}, 054016 (2003);



\bibitem{CG-NPB}C.H. Chen and C.Q. Geng, Nucl. Phys. B{\bf 636},
338 (2002); Phys. Rev. D{ \bf 66}, 014007 (2002).

\bibitem{Aliev} T.M. Aliev {\it
et al.}, Phys. Rev. D {\bf 66}, 115006 (2002).

\bibitem{KO} K. Cheung and O.C.W. Kong, Phys. Rev. D{\bf 64}, 095007 (2001).

\bibitem{CP}E.J. Chun and J.D. Park, JHEP {\bf 0501}, 009 (2005).

\bibitem{Higgs} J.F. Gunion {\it et al.}, {\it The Higgs Hunter's Guide}
(Addison-Wesley, Reading, M.A., 1990).

\bibitem{BurasPLB} A.J. Buras, Phys. Lett. B{\bf 566}, 115 (2003).

\bibitem{D0} D0 Collaboration, V.M. Abazov {\it et al.},
Phys. Rev. Lett. {\bf 94}, 071802 (2005).

\bibitem{BBL}  G. Buchalla, A. J. Buras and M. E. Lautenbacher, Rev. Mod.
 Phys {\bf 68}, 1230 (1996).

\bibitem{CGPRD66}C.H. Chen and C.Q. Geng, Phys. Rev. D{\bf 66}, 034006
(2002).

\bibitem{Ball} P. Ball and R. Zwicky, Phys. Rev. D{\bf 71}, 014015 (2005);
Phys. Rev. D{\bf 71}, 014029 (2005).

\bibitem{FBA} Q.S. Yan {\it et al.},
 Phys. Rev. D{\bf 62}, 094023 (2000) ; D.A. Demir {\it et al.},
Phys. Rev. D{\bf 66}, 034015 (2002);  A. Arda and M. Boz, Phys.
Rev. D{\bf 66}, 075012 (2002); T. Feldmann and J. Ma, JHEP {\bf
0301}, 074 (2003); S. R. Choudhury {\it et al.},
Phys. Rev. D{\bf 69}, 054018 (2004).


\end{thebibliography}
\end{document}